# UNSW-NB15 Computer Security Dataset: Analysis through Visualization

Zeinab Zoghi and Gursel Serpen
Electrical Engineering & Computer Science, University of Toledo, Ohio, USA

## ABSTRACT

This paper presents a visual analysis of the UNSW-NB25 computer network security or intrusion detection dataset in order to detect any issues inherent to this dataset which may require researchers to address before employing this dataset for data-driven model development such as a machine learning classifier. A number of data preprocessing alorithms are applied on the raw data to address common issues such as elimination of redundant features, conversion of nominal features into numerical format and scaling. PCA, t-SNE and K-means clustering algorithms are employed for developing the graphs and plots for visualization. Consequent analysis through visualization identified and illustrated two major problems as class imbalance and class overlap for this dataset. In conclusion, it is necessary to address these two problems of class imbalance and class overlap prior to employing this dataset for any classifier model development.

## 1    INTRODUCTION

The UNSW-NB15 computer network security dataset was released in 2015 (Moustafa & Slay, 2015). This dataset is comprised of 2,540,044 realistic modern normal and abnormal (also known as attack) network activities. These records were gathered by IXIA traffic generator using three virtual servers. Two servers were configured to distribute the normal network traffic and the third one was configured to generate the abnormal network traffic.

A total of 49 features including packet-based and flow-based features were extracted from the raw network packets by Argus and Bro-IDS tools. Packet-based features are extracted from the packet header and its payload (also called packet data). In contrast, flow-based features are generated using the sequencing of packets, from a source to a destination, traveling in the network. The direction, inter-packet length and inter-arrival times are the most important properties in the flow-based feature formulation: total duration (dur) and destination-to-source-time-to-live (dttl) are two examples of flow-based features. The features are categorized into three sets, namely basic (6 to 18), content (19 to 26) and time (27 to 35). Features 36 to 40 and 41 to 47 are labeled as general-purpose features and connection features, respectively. General purpose features category includes those features which are intended to explain the purpose of an individual record while connection features depict the characteristic of the connection among a hundred sequentially ordered records. The last two features include attack categories and labels.

Attacks are categorized as Analysis, Backdoor, DoS, Exploits, Fuzzers, Generic, Reconnaissance, Shellcode and Worms. Normal attacks are represented using 2,218,761 records while Fuzzers, Analysis, Backdoors, DoS, Exploits, Generic, Reconnaissance, Shellcode and Worms signatures include 24246, 2677, 2329, 16535, 44525, 215481, 13987, 1511, and 174 records, respectively. Consequently, there is considerable lack of balance for the dataset as 87% of the dataset comprises Normal records whereas only 0.007% of the dataset consists of Worms records. Developers of the dataset also subsampled and split the dataset into training and testing subsets as presented in Table 1, which has been employed by other researchers (Kanimozhi & Jacob; Kumar, Sinha, Das, Pandey, & Goswami, 2020; Moustafa & Slay, 2016).

The dataset is available on the UNSW web page[1]. The structure of this dataset is more complex in comparison with the other benchmark datasets such as DARPA98 (LABORATORY, 1998), KDDCUP 99 (Janarthanan & Zargari, 2017; Moustafa & Slay, 2016), and NSL-KDD (Tavallaee, Bagheri, Lu, & Ghorbani, 2009) among others. This makes the UNSW-NB15 more comprehensive for evaluating the existing network intrusion detection systems in a more reliable way (Moustafa & Slay, 2016).

**Table 1**. Number of records in training and testing subsets for each class

| Classes | Training Subset | Testing Subset |
|---|---|---|
| Normal | 56,000 | 37,000 |
| Analysis | 2,000 | 677 |
| Backdoor | 1,746 | 583 |
| DoS | 12,264 | 4,089 |
| Exploits | 33,393 | 11,132 |
| Fuzzers | 18,184 | 6,062 |
| Generic | 40,000 | 18,871 |
| Reconnaissance | 10,491 | 3,496 |
| Shellcode | 1,133 | 378 |
| Worms | 130 | 44 |
| Total Number of Records | 175,341 | 82,332 |

Visualization is an essential approach to detect the data formations before a dataset can be meaningfully employed for a data-driven classifier model development such as for intrusion detection. It illustrates the complex data, clearly communicates important information, and offers valuable clues for building an optimized data-analysis model. Although the great amount of research has been done on UNSW-NB15 dataset concerning intrusion detection, several of them (Karami, 2018; Zong, Chow, & Susilo, 2018, 2019, 2020) deployed the visualization techniques prior to designing any classifier models for the data. The research presented in this study aims to provide a visual analysis of UNSW-NB15 dataset to offer a deep insight into the intricacies of the dataset which may result in the data-driven models to demonstrate poor performance. Analysis of the UNSW-NB15 dataset through visual means is expected to expose any problems that may hinder the performance of classifier models.

---

We implemented several preprocessing techniques to prepare the data for the visual analysis of data. We first check if there are any redundancy in UNSW-NB15, transform the nominal input features to numerical, rescale them, and select the relevant input features. We utilize PCA in order to project the preprocessed data into low-dimensional space. Finally, 3D scatter plots, t-SNE, and K-means intercluster distance map are the visualization techniques employed in this study.

## 2    LITERATURE REVIEW

A principal objective of information visualization is to translate large and complex datasets and illustrate them in a visual format. This process facilitates data interpretation in order to identify patterns, trends, clusters, correlations and relationships among the data points. There are many different techniques and tools to visualize data. Some prominent data visualization techniques are weight table, bar chart, heat map, box plot and scatter plot.

Many data visualization techniques were reported for a variety of domains in the recent literature. Stahnke et al. (Stahnke, Dörk, Müller, & Thom, 2015) employed Multidimentional Scaling (MDS) to project high-dimentional data into a low-dimensional space and visualize the data using scatterplot and heatmap. The purpose of using MDS was to find the distance dissimalirites for any pair of datapoints. They proposed the concept of 'probing' that represents an integrated method containing dynamic selections, class selection and clustering, providing a valuable insight into data and correct the projection errors. They demonstrated the application of their method on the dataset of OECD countries.

Wine Recognition dataset was utilized to evaluate the performance of c-PCA visualization method proposed by (Fujiwara, Kwon, & Ma, 2019). The authors used t-distributed Stochastic Neighbor Embedding (t-SNE) (Maaten & Hinton, 2008) to visualize high-dimensional data. Three clusters were generated by applying Density-Based Spatial Clustering of Applications with Noise (DBSCAN) (Ester, Kriegel, Sander, & Xu, 1996) to the output of t-SNE . At the final step, c-PCA was implemented on the clusters and the engagement rate of features in each cluster was calculated. The engagement rate identified the disimilaries between one cluster and the rest.

İn (Kwon et al., 2018), the authors proposed a deep learning model using a recurrent neural network (RNN) called RetainEX. The model was implemented on electronic medical records containing the events such as histories of patients' diagnoses and medications. This method is comprised of visual analytics carried out by RetainVis visualization tool. To form a 2D shape of patients' medical information, t-SNE was implemented on contribution scores calculated from all codes that each patient had for every visit. The final 2D projection view of Electronic Medical Records (EMRs) data was summarized by representative points using clustering approaches. Similarly, (van Unen et al., 2017) implemented dimentionality reduction-based techniques on medical datasets. They utilized Hierarchical Stochastic Neighbor Embedding (HSNE) in order to eliminate the scalibility limit of t-SNE by considering rare cell populations that were removed in the last undersampling procedure and reproduce former observations.

Wilkinson (Wilkinson, 2017) presented a new algorithm called HDoutliers to visualize the observations that are remarkably far from the closest point to the majority number of the datapoints in high-dimensional space. The algorithm brings all the variables to the same range to prevent disproportionate impact on Euclidean distances due to data skewness. The authors employed the Leader algorithm (Hartigan, 1975) to characterize high density of datapoints using a large number of small balls rather than small number of clusters. The distance of the members in each cluster was calculated using the nearest neighbor algorithm. The values were deployed for outlier identification or anomaly detection procedure.

Data transformation, pixel-based representation, graph representation, and coordinated multi-views (CMV) are four principal visualization techniques employed in (Ji, Jeong, & Jeong, 2020). Principle Component Analysis (PCA), Discrete Wavelet Transform (DWT), and Multi-Dimensional Scaling (MDS) were utilized to visualize the degree of disimilarities that may exist in the data points of a large dataset. They were shown in low-dimensional space, based on pairwise distance. İn the pixel-based representation method, datapoints were mapped in colored pixels using space-filling design technique. Also, the authors used node-link diagram to convert network traffic to a graph consisting of nodes and links. The graphs were illustrated using PCA, hierarchical clustering, and force-directed graph drawing algorithm. The CMV framework in this study provided multiple views of a data. The results of PCA computed from the network traffic were shown using this framework designing with the parallel coordinates visualization technique. The relationship of input features were depicted using scatter plot during the final stage.

Ruan et al. (Ruan, Miao, Pan, Patterson, & Zhang, 2017), deployed hash algorithm, weight table, and sampling method to address volume, variety, and velocity issues arising from big data. They formed a weight table and assigned large weights to the classes with smaller population to garantee the less probable data records to be selected by the sampling method. The data records were selected based on the weight table and the hash code retrieved during the sampling approach which minimizes class imbalance issues and any redundancies, respectively. MDS and PCA were used for big data visualization on the resampling data. Karami (Karami, 2018) proposed a modified Self-Organizing Map (SOM) to both detect the anomalies and provide the information summary from 2D data visualization using UNSW-NB15. Similarly, (Zong et al., 2018, 2019, 2020) visualized the UNSW-NB15 datapoints. They utilized PCA to reduce the dimentionality and keep the first three principal components. As for the visualization, they employed a voxel-based approach to visualize the ML decision space. They calculated 3D coordinate as a weighted sum of the first 3 components and multiplied the results by a constant to control the distance between the voxels.

## 2     PREPROCESSING

In this section, we first verify the existence of redundant input features in UNSW-NB15 data records, convert the existing nominal input features to numerical format, map them into the same scale, and extract the most informative input features to prevent the features from misleading clustering results as well as reducing the computational cost in the visualization phase (Cao, Zhao, & Zaiane, 2013; Cieslak & Chawla, 2008; Imam, Ting, & Kamruzzaman, 2006; Köknar-Tezel & Latecki, 2009).

We utilized the training and testing subsets of UNSW-NB15 as provided by Moustafa et al. (Moustafa, 2017). They selected 175,341 records to form the training subset and 82,332 records for the testing subset among the original 2,218,761 records. We employ the training subset for data visualization. This dataset contains 49 features. However, not all 49 features are necessarily relevant for the class labels. Two input features, namely record_start_time and record_last_time, are redundant due to the presence of total duration (dur) that is obtained from the difference of values for these two features. Since some features are specific to the computing infrastructure such as source IP address, source port number, destination IP address, and destination port number, they do not possess relevant information for intrusion detection purposes. Accordingly, we eliminate record_start_time, record_last_time, source IP address, source port number, destination IP address, and destination IP address and keep the explanatory input features by some prior knowledge gained by the class feature. For the 43 remaining input features, two of them are class features, namely attack_cat and label. The attack_cat feature is of type nominal and contains the names of attack categories. For visualization, we need this feature to present the name of the overlapping and imbalanced classes. Also, for the purpose of multi-class classification, this feature is needed. The feature (class) label is binary valued for which a 0 value indicates normal and a 1 value shows attack records, which is relevant for binary classification cases.

Among the 41 non-target features, three of them are nominal. We convert the nominal features to numerical as most of the machine learning models and scalers can readily work with the numerical values. To convert the nominal to numerical, we use label encoder implemented as in scikit-learn in Python.

In order to map the input features to the same scale, six different data transformation algorithms called data normalizer, min-max scaler, robust scaler, standard scaler, quantile transformer, and power transformer were implemented and evaluated for suitability on the numerical input features. We measure nearest shrunken centroid (Tibshirani, Hastie, Narasimhan, & Chu, 2002) for each attack class as well as normal class type and calculate the distance between class centroids using Mahalanobis distance (Mahalanobis, 1936; ur Rehman, Khan, & Naveed, 2019) before and after applying the scalers. The results are illustrated through seven heatmaps. The heatmap made for the original dataset in Figure 1 is compared with the six other heatmaps in Figure 2 representing the distances between class centroids when the scalers are applied. The distances of the centroids are depicted with the range of colors between red and green. A red single cell in a heatmap indicates that the centroids of the corresponding classes are closer and a green cell represents that the centroids of the corresponding classes are far from each other. Comparison of heatmaps in Figure 2 shows that min-max scaler shifts the values to the area where the class centroids are placed in the farther points against each other while the data points belonging to one attack category gather around in the same area. This property of min-max scaler helps to present the data with less overlap yet higher clarity, and potentially improve the performance of the machine learning model in the subsequent classification phase. However, the class overlap issue is partially addressed in this step. It will be observable in the subsequent graphs and plots that the dataset still has the class overlap although to a lesser degree.

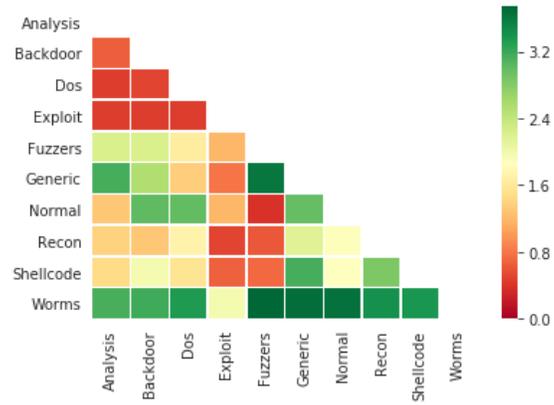

**Figure 1** The Mahalanobis distance of the centroids when the dataset in not normalized

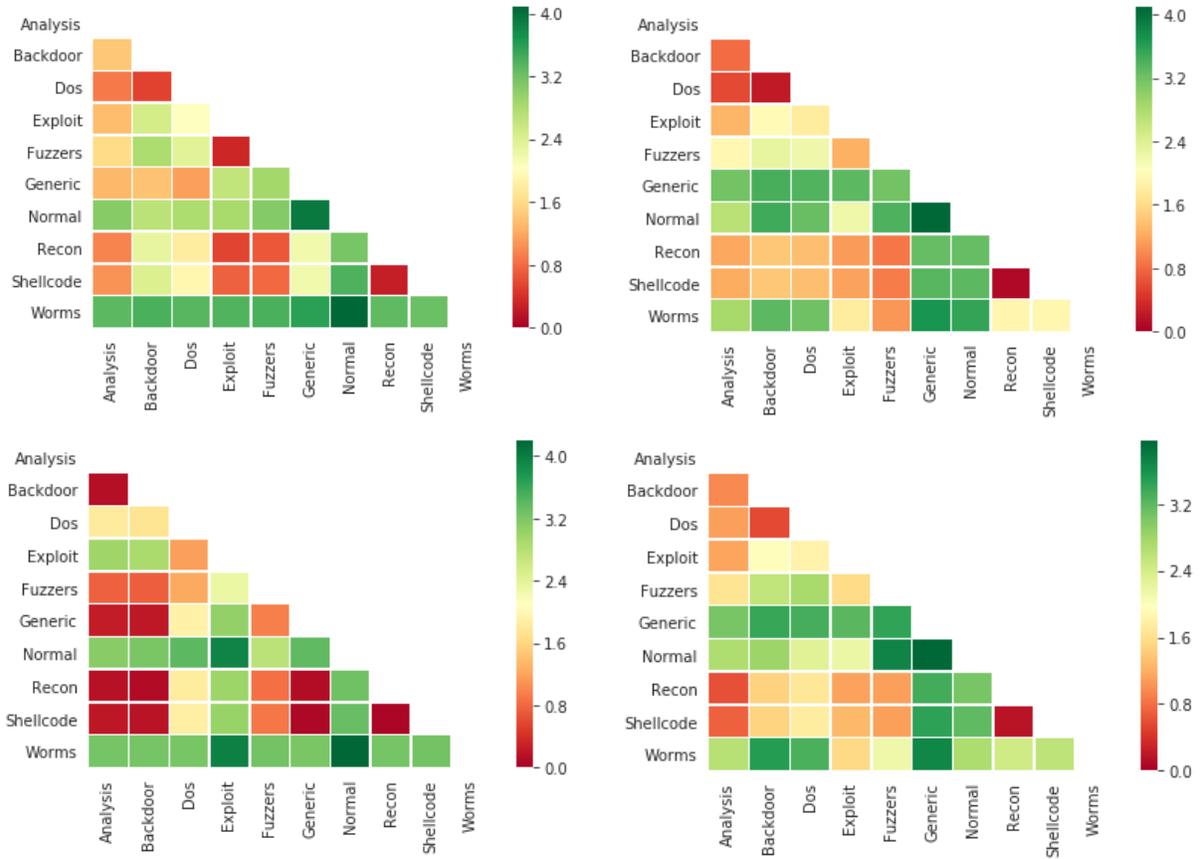

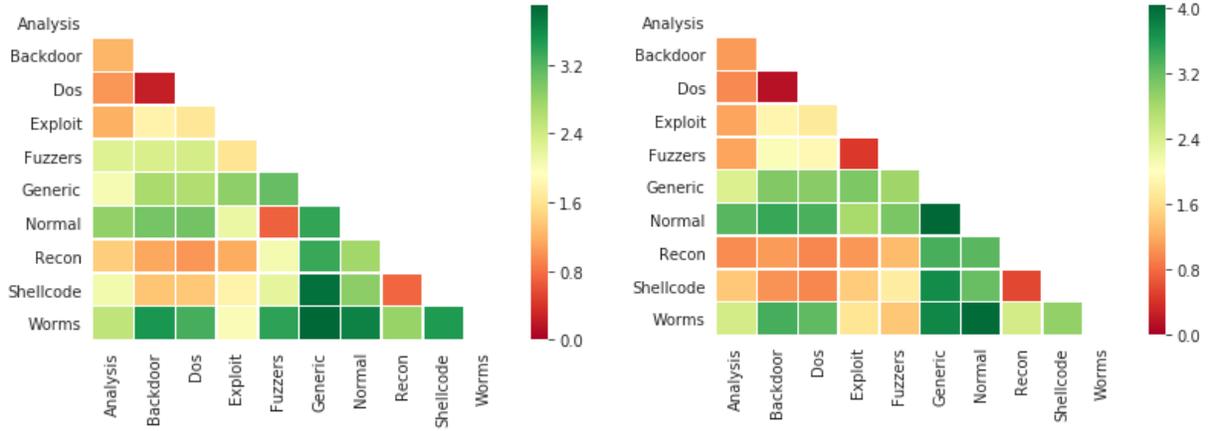

**Figure 2** The Mahalanobis distance of the centroids after normalizing the data (upper-left), scaling the dataset with min-max, maxabs (upper-right), robust (left-center), standard (right-center), quantile transformation (lower-left), and power transformation (lower-right)

We implement Elastic Net algorithm (Zou & Hastie, 2005) on the dataset processed by min-max scaler in order to reduce the number of input features and subsequently decrease the computational cost associated with the process of visualization and also to improve the performance of the machine learning models. This algorithm is applied 5 times using 5-fold cross validation with different values for alpha determined randomly and and L1 ratio with the value 0.5, by the ElasticNetCV implementation in Python. The mean squared error (MSE) and regularization path are calculated and shown in Figures 3 and 4, respectively. MSE is calculated as

$$\frac{1}{n}\sum_{i=1}^{n}(Y_i - \hat{Y}_i)^2$$

where $n$ is the number of samples; $Y$ is the vector of real class or target values; and $\hat{Y}$ is the vector of predicted class or target values. In Figure 3, 5-fold cross validation which is also known as cyclic coordinate descent, is applied across all the alpha values from 0 to 0.001 to find the minimum alpha and solve Elastic Net penalized regression model. As alpha value gets larger, the mean squared error in each fold also increases. The minimum amount of error is reported at the point where the optimum alpha value is roughly 0.0003. Accordingly, we also select 0.0003 as the alpha value to train the Elastic Net algorithm in order to eliminate the less informative features for which the mean squared error is also minimum.

In order to implement the features selection based on the optimum alpha and subsequently improve the data interpretation in the visualization phase, we generate the regularization paths. We find the irrelevant features and eliminate them where the alpha is 0.0003 using the Elastic Net algorithm. This algorithm zeroes out the coefficients of irrelevant features. Figure 4 presents the regularization paths. lambda is the number of alphas used for L1 ratio or the list of L1 ratios along the regularization path. In this case, we chose 100 for this variable. The greater the value of this variable is, the smoother the curves that are shown in the graph are. We negate the log base 10 of lambda to reverse the graph and interpret it from right to left as it should. There are 41 curves in this figure, some are overlapping with each other. Each curve represents an input feature. From right to left in this graph, by reaching to the point 0.0003 where alpha was minimum, 17 curves touch the coefficient boundary which is zero. These features depict high correlation and removing them should not affect the performance of the classifier models. In other words, removing these features helps to improve the performance of the final machine learning model and decrease the computational costs in the visualization phase. Therefore, these features are eliminated. The entire process of inputting the Elastic Net algorithm with the training subset to selecting a value for alpha takes approximately 4.70 seconds where the algorithm converges at the 409th iteration (on a platform with Windows 7 Enterprise 64-bit operating system, Intel® Core™ i5-4690 CPU @ 3.50GHz processor, and 16.0 GB RAM). The optimal alpha value chosen by cross validation in the previous paragraph is used for feature selection approach and 17 non-informative features are eliminated. Consequently, 24 most informative and the least correlated features among the 41 input features are selected in this phase as presented in Table 2.

**Table 2**. The list of features selected by Elastic Net

| Feature No. | Input Feature Name | Discription |
|:-----------:|:-------------------|:------------|
| 1 | dur | Record total duration |
| 2 | proto | Transaction protocol |
| 3 | service | Contains the network services |
| 4 | state | Contains the state and its dependent protocol |
| 5 | spkts | Source to destination packet count |
| 6 | dpkts | Destination to source packet count |
| 7 | sbytes | Source to destination transaction bytes |
| 8 | dbytes | Destination to source transaction bytes |
| 9 | rate | Ethernet data rates transmitted and received |
| 10 | sttl | Source to destination time to live value |
| 11 | dttl | Destination to source time to live value |
| 12 | sload | Source bits per second |
| 13 | dload | Destination bits per second |
| 14 | sloss | Source packets retransmitted or dropped |
| 15 | dloss | Destination packets retransmitted or dropped |
| 16 | sinpkt | Source interpacket arrival time (mSec) |
| 17 | dinpkt | Destination interpacket arrival time (mSec) |
| 18 | sjit | Source jitter (mSec) |
| 19 | djit | Destination jitter (mSec) |
| 20 | swin | Source TCP window advertisement value |
| 21 | stcpb | Destination TCP window advertisement value |
| 22 | dtcpb | Destination TCP base sequence number |
| 23 | dwin | Destination TCP window advertisement value |
| 24 | tcprtt | TCP connection setup round-trip time |

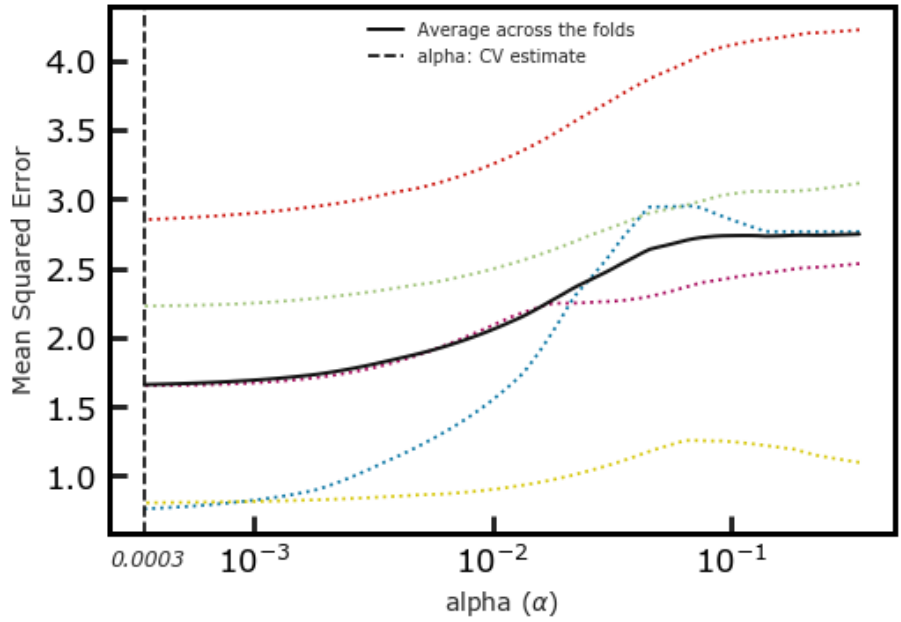

**Figure 3**. The mean squared error calculated for Elastic Net through 5-fold cross validation

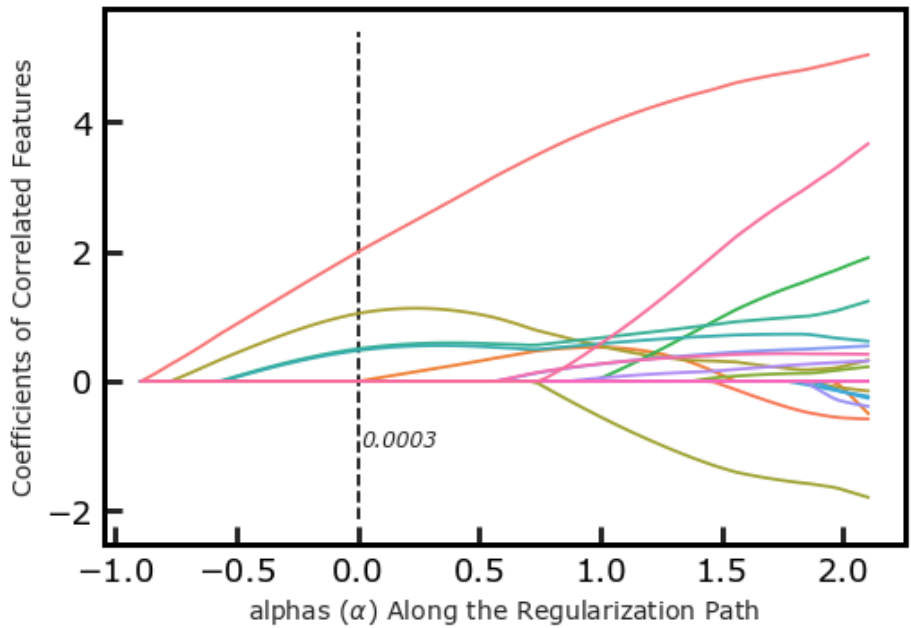

**Figure 4**. The regularization path for Elastic Net presenting the informative features

### 3    VISUALIZATION

The data visualization analysis is conducted on Windows 7 Enterprise 64-bit operating system, Intel® Core™ i5-4690 CPU @ 3.50GHz processor, and 16.0 GB RAM. To avoid the clutter issue (Kwon et al., 2017) in the visualization phase, which may arise due to a large number of data points, we utilized limited number of the data points to present the data using t-SNE and K-means clustering where the data points are illustrated in 2D projection view. We deployed stratified sampling (Taverniers & Tartakovsky, 2020) to randomly extract 20% of the records from each class type. These records are randomly selected from the training subset and forms a new subset containing a smaller number of records to present UNSW-NB15 with more clarity.

Two prominent problems are identified through this analysis. The issues are known as class imbalance and class overlap. Class imbalance is composed of between-class and within-class imbalance. Between-class imbalance corresponds to the case where one class or multiple classes in a dataset are underrepresented in comparison with other classes. In other words, a dataset shows significantly unequal distribution among its classes. To illustrate this problem, we plot the distribution of entire attack classes for the UNSW-NB15 dataset in Figure 5. As seen in this figure, normal records constitute 87% of all records while the combined record count for all 9 attack classes is only 13%. Additionally, 13% of records in the overall dataset are not equally distributed among the 9 attack classes as 65% of all attack records belong to the Generic attack class while only 0.0008% of all attack records belong to the Worms attack class.

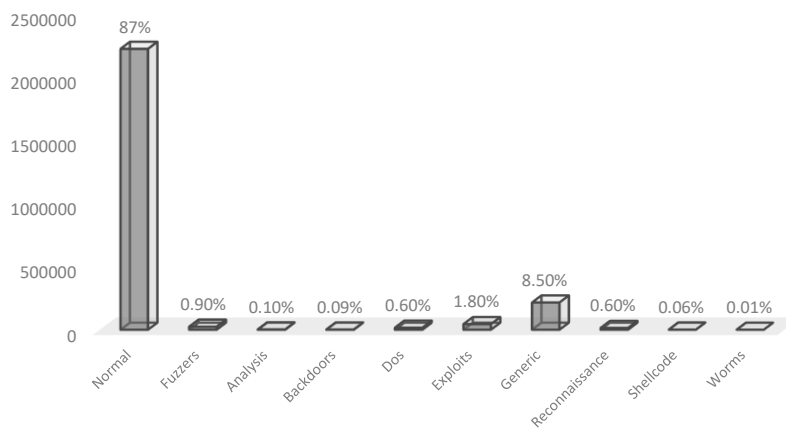

**Figure 5.** Class Distribution of UNSW-NB15 Dataset

Within-class imbalance, on the other hand, represents the case where one class is comprised of several different subclasses with different distributions. To discover whether the classes are made up of imbalanced sub-clusters, we use two visualizing techniques from scikit-learn machine learning library in Python, namely PCA with the time complexity of $O[\min(n^3, p^3)]$ (Johnstone & Lu, 2009) where $n$ is the number of data records and $p$ is the number of input features and t-SNE with time complexity of $O(n^2)$ (Pezzotti et al., 2016). PCA is deployed to project the data to 2 and 3 dimensions before presenting the data using the scatter plot, t-SNE, and K-means clustering. It took PCA and t-SNE to generate Figures 6 to 8 in 1.8 sec and 2178.33 sec, respectively. Figures 6 and 7 show the within-class imbalance in the UNSW-NB15 dataset. The class types appear in different colors tagged with the name of class types on the plot legend added to the right side of the plots. As Figure 6 shows, the Exploits attack class is composed of several different size clusters where clusters are compact. The clustering is different for the Worms attack class for which there are two, but different size cluster groupings as shown in Figure 7. The larger cluster grouping is composed of multiple subgroupings with gaps in between.

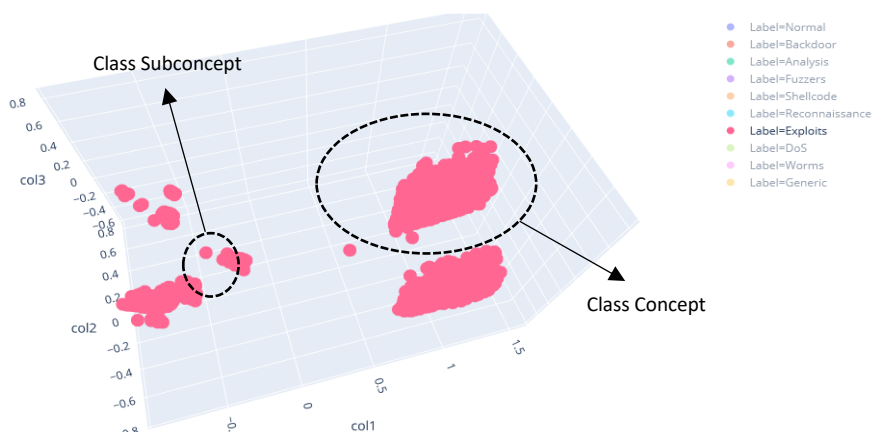

**Figure 6.** Visualization of the Exploits instances using PCA

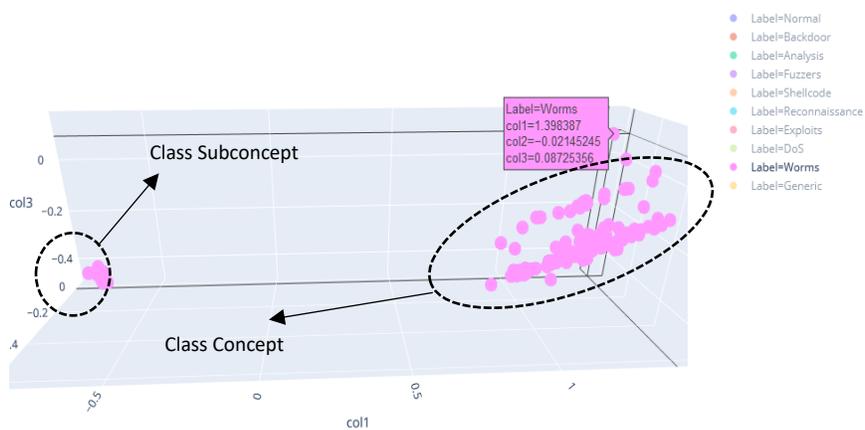

**Figure 7.** Visualization of the Worms instances using PCA

Figure 8 illustrates all the concepts using a 2-dimensional scatter plot for the data points in the dataset. The class types have multiple clusters of different sizes and spread across the two-dimensional analysis space. Many classes are composed of a few relatively large clusters and many small clusters. Additionally, the boundaries separating classes are not clear cut: there is noticeable overlap between or among multiple clusters belonging to different classes. For this method, Kullback-Leibler (KL) divergence score is calculated from 250 iterations at the early stages of optimization to 1000 iterations. The divergence score varied from 96.52 to 2.93 which represents the measure of how one probability distribution defers from the second. During the 250 iterations, 91 nearest neighbors of 175,341 samples are computed in 179.02 sec, the samples are indexed in 39.35 sec, and conditional probability assessments and KL divergence occurred in 1959.96 sec.

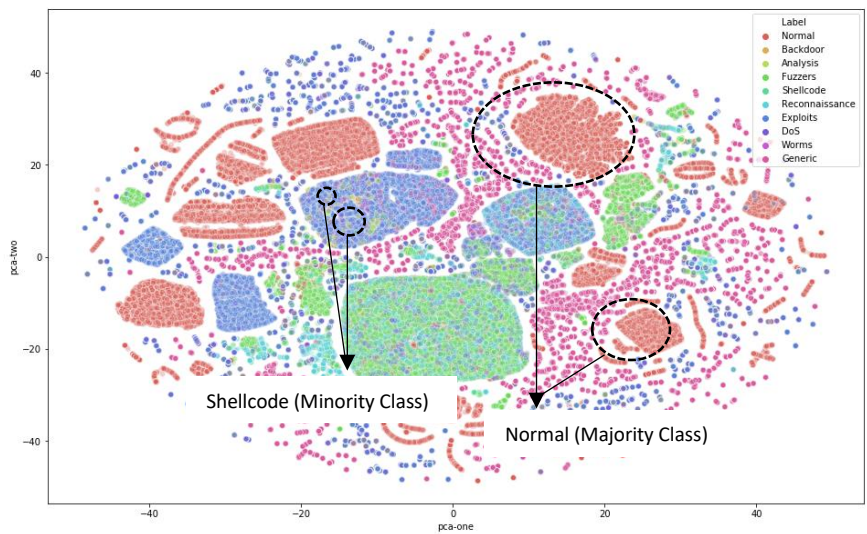

**Figure 8.** Visualization of the entire dataset using t-SNE

This dataset also suffers from the so-called "overlap problem." Many attack class records mimic the behavior of the Normal records. While the emphasis of intrusion detection systems is detecting and/or identifying the malicious network traffic, if it is trained by this dataset without addressing the overlap problem, a satisfactory outcome will not be achieved. In order to expose the degree or scale of this problem, we first sketch the data points in a 3-dimensional scatter plot as shown in Figure 9. The same figure shows that many attack classes overlap as indicated with the content of the dotted circles.

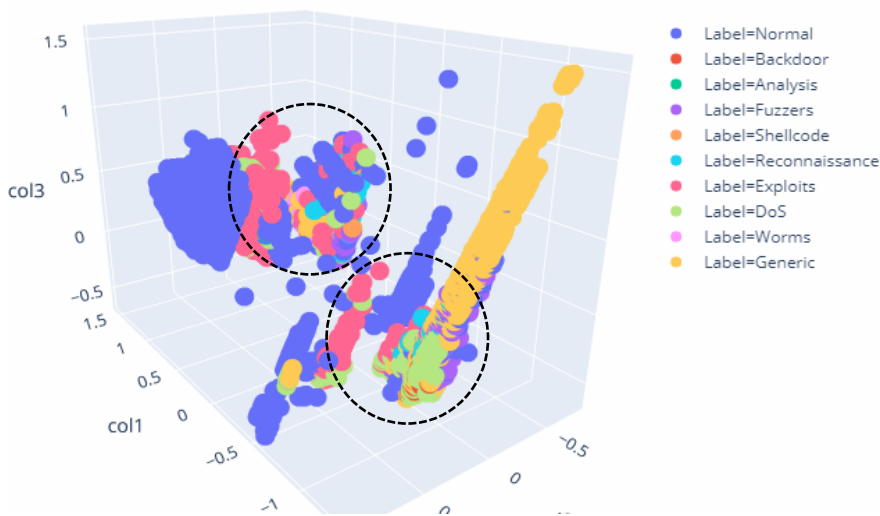

**Figure 9.** Visualization of the entire dataset using PCA for class overlap

We further investigate for those attack class records which reside in the subspace where primarily Normal records are present. Figures 10 and 11 detail the overlap cases among a subset of attack classes and the Normal class. As Figure 10 illustrates that Exploits attack class have the overlapping problem with the Normal class. In Figure 10, there is noticeable overlap between the sets of data points belonging to Exploits (in red) and those belonging to Normal class (in navy blue). In Figure 11, it is easy to observe that records belonging to Fuzzers and Normal classes are clustered together for the cluster in the lower left.

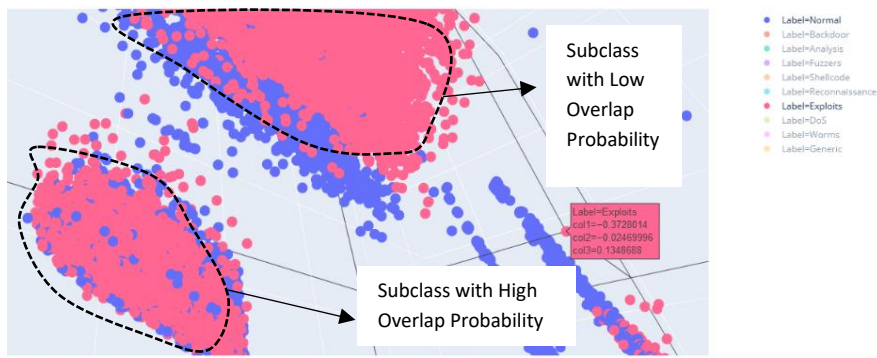

**Figure 10.** Visualization of the Normal and Exploits data points

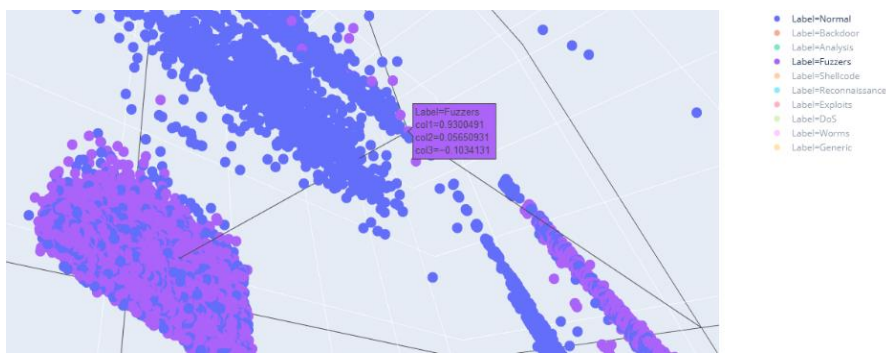

**Figure 11.** Visualization of the Normal and Fuzzers data points

Figure 12 shows overlaps for the entire dataset. We used the *K*-means clustering method to compute the distance between the class centroids that was already calculated using nearest shrunken centroid algorithm. Inter-cluster distances are also utilized to sketch the map. Intercluster distance maps illustrates an embedding of the class centers in 2D view with the distance against other centers. The closer the circles drawn in the map, the closer the data points are in the original feature space. As the figure shows, the degree or amount of overlap among classes 3, 4 and 7 as well as between 0 and 6, and between 1 and 2 are substantial.

We employ K-means clustering with time complexity of $O(n^3)$ along with InterclusterDistance visualizer in Python to present the distance between the clusters (Xiang, Zhao, Li, Hao, & Li, 2018). It takes 0.001 sec for these algorithms to generate Figure 12.

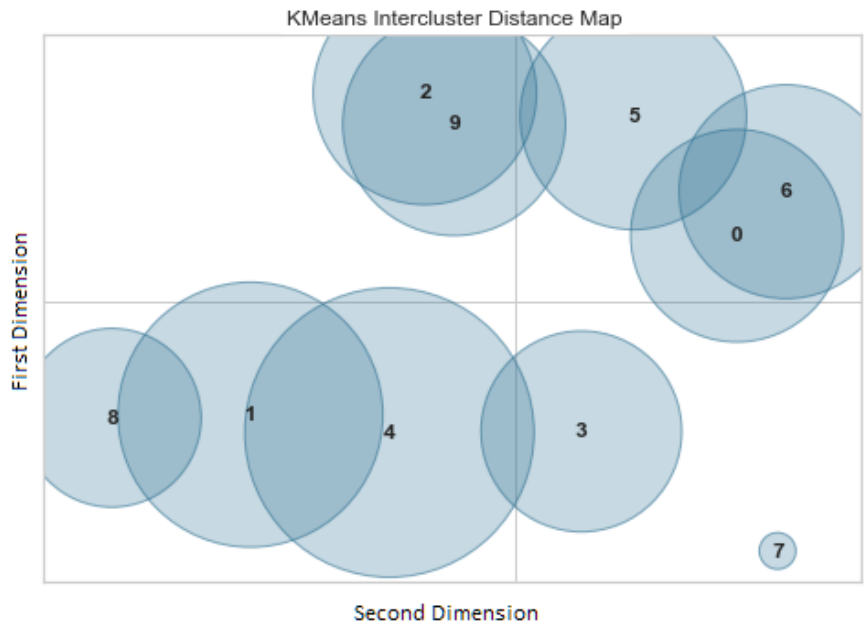

**Figure 12.** Illustration of class overlap using inter-cluster distances and K-means

The clusters are sized according to the number of the members in a single cluster. In other words, the number of the samples that their distance to the centroid are much shorter than the other data points belong to the corresponding center that forms a cluster. The map gives a sense of the number of data points in a cluster, how they spread all across the cluster, and the probability of class overlap.

## 4    CONCLUSIONS

The UNSW-NB15 dataset facilitates empirical studies on intrusion detection system development using data-driven approaches. However, this dataset has two major issues, namely class imbalance and class overlap, which need to be addressed prior to being employed for model development. Class imbalance and overlap, if not addressed, are likely to hinder the attack detection and identification performance of intrusion detection systems. In preparation for the subsequent visual analysis, several preprocessing steps were implemented which included removing redundant features, normalizing the features, and scaling the features. Following the data preprocessing we used a variety of visualization techniques to expose and illustrate class imbalance and class overlap problems with this dataset. The dataset was projected into 2D and 3D views using PCA. It was then analyzed with the scatter plot, t-SNE, and K-means inter-cluster distance map. Given the degree of severity as presented through the visualization it is imperative that effective approaches need to be implemented to mitigate the adverse effects of these two problems on the classification performance of any data-driven statistical or machine learning model.